# Relative Sea Level and Abrupt Mass Unloading

**Julien Gargani[1,2]**

[1]Université Paris-Saclay, CNRS, Geops, Orsay, France.

[2]Université Paris-Saclay, Centre d'Alembert, Orsay, France.

**Highlight**

- An abrupt mass unloading of 30 ± 10 km$^3$ occurred in Barbados and caused isostatic adjustment of 0.45 ± 0.15 mm/yr.
- An uplift increases from 0.34 to 0.8 mm/yr at 11.2 ± 0.1 kyr BP in less than 0.2 kyr explain coral reef elevation in Barbados.
- 150 yr after the termination of Younger Drias and 300 yr before the abrupt mass unloading, a sea-level jump of 4.8 m occurred, corresponding to meltwater pulse MWP-1B.

**Abstract**

Relative sea level records climatic change as well as vertical land movement. In Barbados, uplift variation is necessary to interpret one of the most complete coral reef records. Here we show that an abrupt mass unloading of 30 ± 10 km$^3$ caused an uplift variation of ~0.45 ± 0.15 mm/yr using a modelling approach. Simulations have been conducted for different volumes and elastic thicknesses. Isostatic adjustment in relation with an abrupt mass unloading explains the observed uplift rate increased from 0.34 mm/yr to 0.8 mm/yr that occurred 11.2 kyr ago. The reconstructed sea-level curve highlights a sea-level jump of 4.8 m, with a delay of 150 yr from the termination of Younger Dryas cold event and 300 yr before the abrupt mass unloading. This sea-level jump corresponds to meltwater pulse MWP-1B and is not an artefact. A stagnation of 500 yr occurred from 12 to 11.5 kyr BP. Relative sea level records are useful to detect past landslides and erosion. Accurate analysis and reconstruction of sea-level permits to determine sea-level abrupt rise caused by climate warming during the last thousand years.

**Keywords :** uplift; climate; coral reef; isostasy; meltwater pulse; landslide

## 1 Introduction

Rapid sea-level rise could be better understood studying ongoing changes in the climate system (Edwards et al., 2021; Hoojer and Vernimmen, 2021) and by studying the abrupt sea-level changes that occurred in the past (Gargani and Rigollet, 2007; Bard et al., 2010; Bache et al., 2015). Reconstruction of sea level could be based on well log and stratigraphic analysis (Haq and Schutter, 2008), seismic and geomorphological observations (Bache et al., 2015) or coral reef studies (Mesolella, 1967; Fairbank, 1989).

Reconstruction of sea level from coral reef must take into account tectonic movement (Radtke and Schellmann, 2006; Peltier and Fairbank, 2006), glacio-hydro isostasy (Austermann et al., 2013) and uncertainty associated with coral reefs species (Bard et al., 2016). Isostatic adjustment could be also caused by erosion (Gargani, 2004). However, coastal erosion is often underestimated (Regard et al., 2022) and consequences on isostatic adjustment are unknown. Furthermore, numerous coastal and marine landslides occur (Urgeles and Camerlinghi, 2013) and could cause isostatic adjustment (Smith and

Wessel, 2000). Unloading must be taken into account to interpret vertical land movement and for accurate reconstruction of relative sea level curves.

Accurate reconstruction of relative sea level curves is of fundamental importance due to the numerous applications and implications. Relative sea level reconstruction could be used to determine local paleogeography (Zinke et al., 2003), to estimate regional vertical land movements by comparing with a reference's sea level curve (Camoin et al., 2004; Pedoja et al., 2014) and to discuss paleoclimate variation (Fairbank, 1989; Blanchon and Shaw, 1995).

One of the most complete records of post-glacial sea-level is based on coral reef drilled at Barbados. Coral reef records in Barbados have been studied for several decades (Mesolella et al., 1969; Bender et al., 1979). These records were used to calibrate relative sea level during the last 500 kyr (Bard et al., 1990; Bard et al., 1990b; Gallup et al., 1994; Peltier and Fairbank, 2006; Abdul et al., 2016) as well as to determine local uplift (Radtke and Schellmann, 2006).

In this study, the role of an abrupt mass unloading in the Tertiary terrain of the Scotland District in the northeast of Barbados is investigated to estimate the potential uplift change of coral reef located in the southwest of the island, near Bridgetown (Figure 1), from 14 to 9 kyr Before Present (BP). More fundamentally, the new sea level reconstruction proposed in this study will be used to quantify the meltwater pulse MWP-1B (age, duration, amplitude).

## 2 Geological context

Barbados Island (430 km$^2$) lies on an accretionary prism above the subducting Atlantic lithosphere that passes beneath the crust of the Caribbean plate (Westbrook et al., 1988)(Figure 1). Deformed Tertiary sedimentary rocks are located below an uplifted Quaternary reef-related cover (Speed, 1981) with a maximum age of 500-600 kyrs (Radtke and Schellmann, 2006). The cause of passive uplift of sedimentary layers could be due to compressive tectonic at the plate junction, mud diapirism or shale movements in deep overpressured zones (Deville et al., 2006).

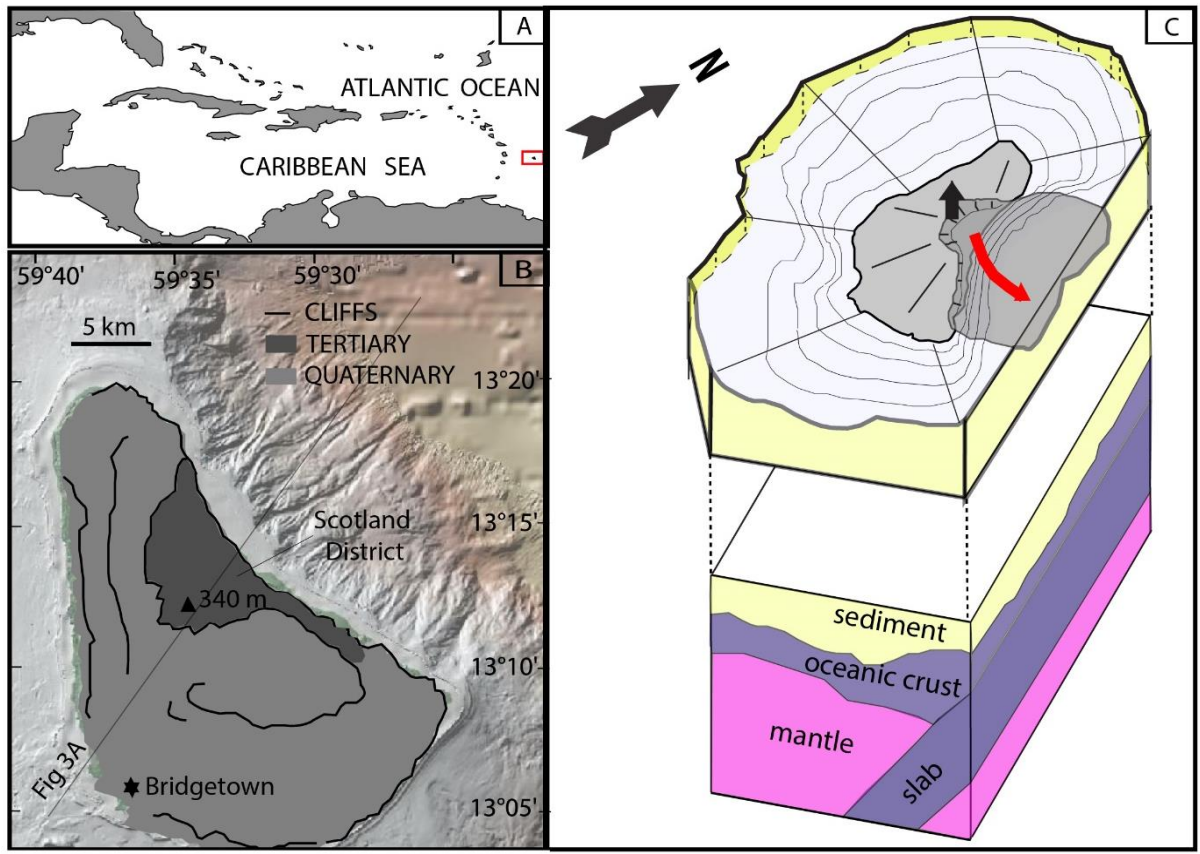

**FIGURE 1: Geographic and geological context. (A) Location of Barbados Island in the red square, (B) Geological map and bathymetry of Barbados Island, (C) Schematic representation of mass unloading and lithosphere geometry. Bathymetry from https://www.ncei.noaa.gov/maps/bathymetry.**

Uplift rate in Barbados was not constant during the last 500 kyr (Radtke and Schellmann, 2006) (Figure 2). From 200 to 500 kyr, uplift rate was around 0.2 mm/yr whereas from 100 to 120 kyr ago, mean uplift rate was estimated to be of $0.47 \pm 0.02$ mm/yr (Radtke and Schellmann, 2006). From 35 to 11.2 kyr BP, an uplift rate of 0.34 mm/yr is often assumed (Peltier and Fairbank, 2006; Abdul et al., 2016). This uplift rate of 0.34 mm/yr was caused by tectonic compression or diapirism (Speed, 1981; Westbrook et al., 1988; Bard et al., 2016).

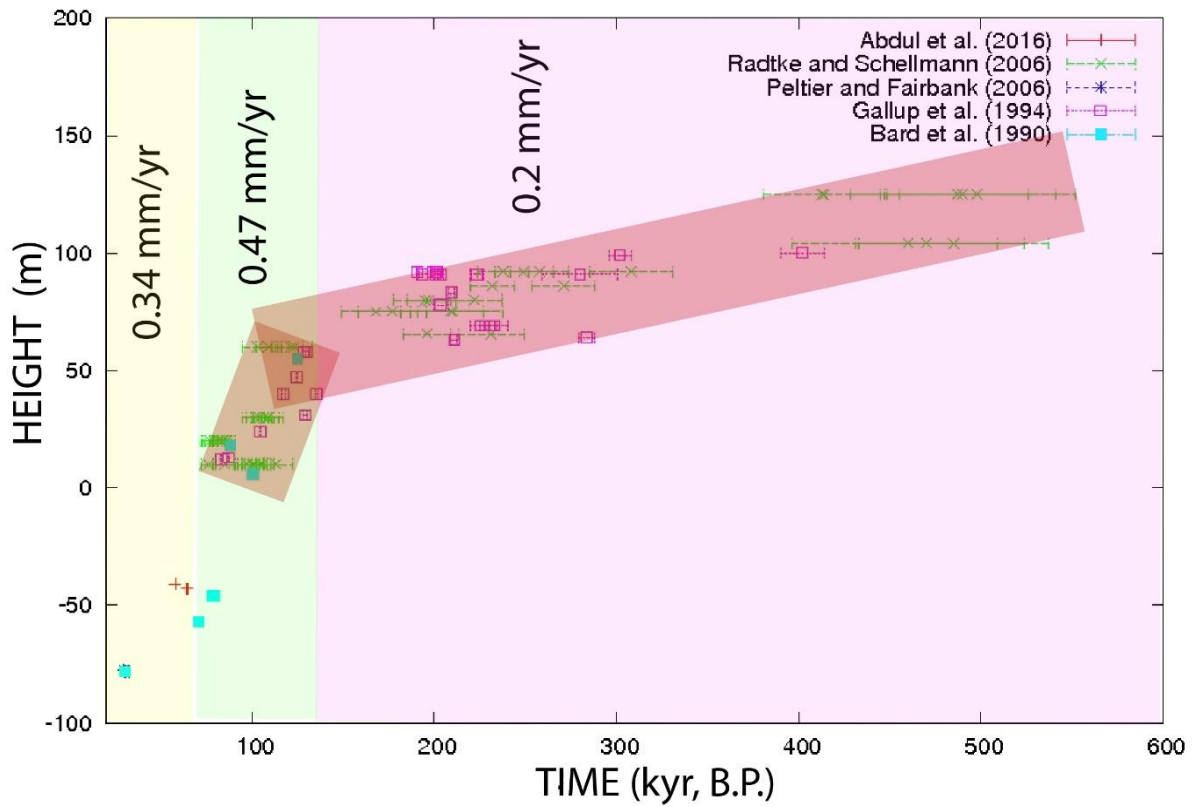

**FIGURE 2: Coral reef elevation and uplift. (A)** Height of coral reefs in Barbados with ages ranging from 20 to 600 kyrs (Bard *et al.*, 1990b; Gallup *et al.*, 1994; Peltier and Fairbank, 2006; Radtke and Schellmann, 2006; Abdul *et al.*, 2016) and uplift rates.

Without taking into account uplift correction, Barbados coral reefs have a significant offset from sea-level reconstruction suggesting the existence of an uplift (Peltier and Fairbank, 2006; Radtke and Schellmann, 2006; Abdul et al., 2016). The observed vertical shift from sea-level curve range from $10 \pm 2$ m at 10.5 kyr to $4 \pm 1$ m at 13 kyr (Figure 3). At 11.2 kyr BP, there is less data during around 200 yr.

Recently, it was highlighted that there were several meters of offset between coral reefs younger than 11.2 kyr and those older than 11.2 kyr (Abdul et al., 2016). This variation of altitude was first attributed to a meltwater pulse record (MWP-1B) (Bard et al., 1996; Abdul et al., 2016), but this offset was discarded by comparing with sea-level curves established in other places (Carlson and Clark, 2012; Lambeck et al., 2014; Bard et al., 2016). This shift was believed to occur 0.4 kyr after the end of Young Dryas (Abdul et al., 2016) that finished $11.610 \pm 0.040$ kyr BP (Cheng et al., 2020). Nevertheless, the possibility that an uplift variation occurred with time from 14 to 9 kyr has not been studied in Barbados, in spite of older uplift variations (Radtke and Schellmann, 2006).

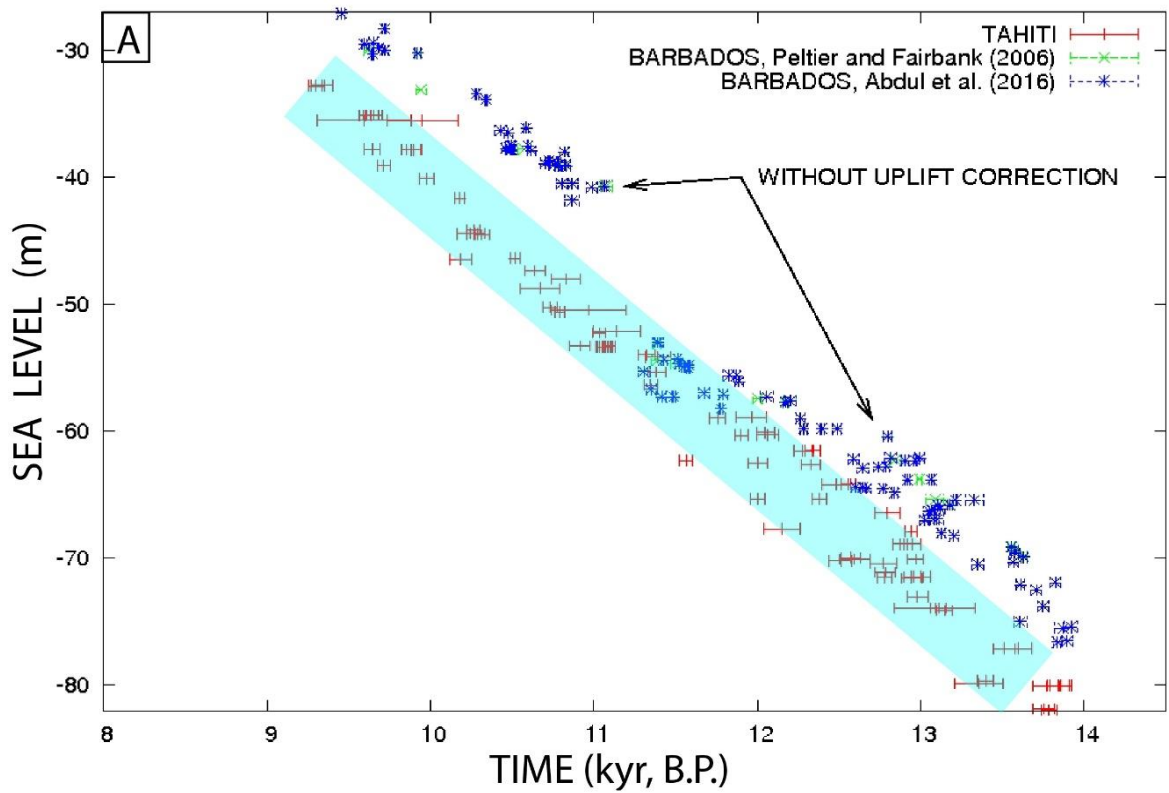

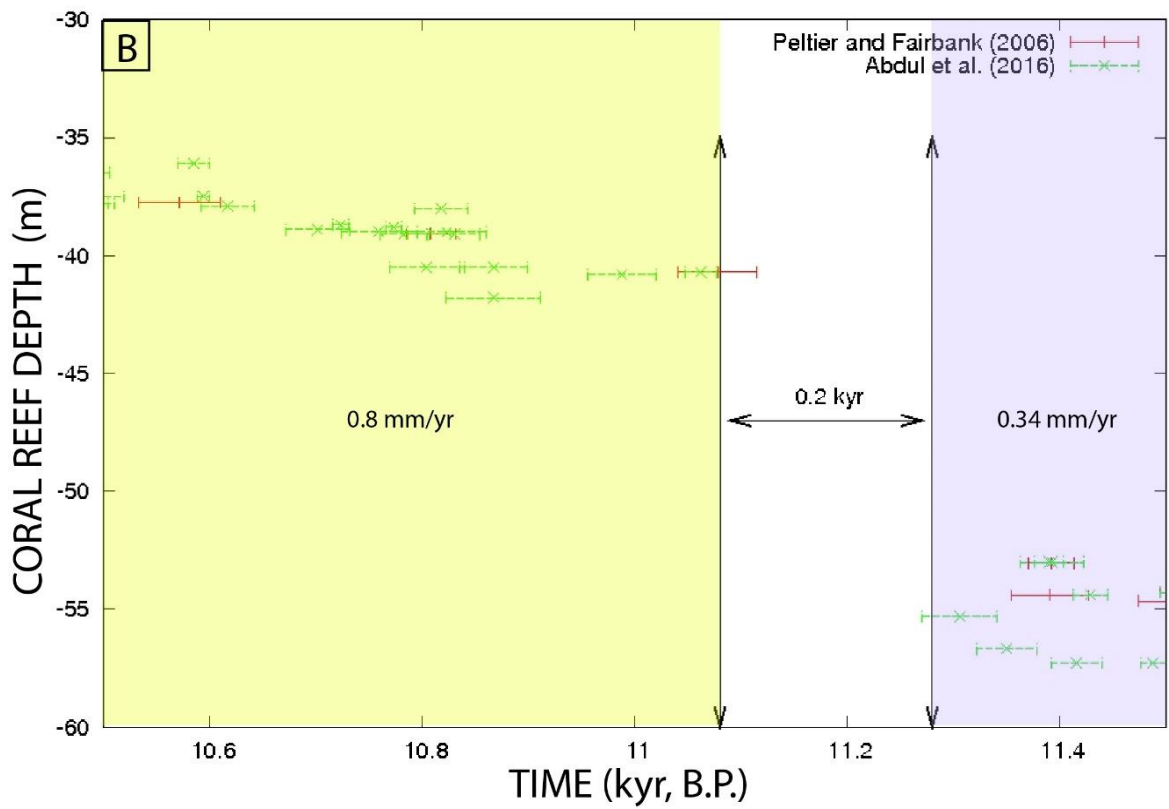

**FIGURE 3:** Coral reef elevation and uplift. (A) Sea level estimated from Tahiti coral reefs corrected from subsidence (0.25 mm/yr) in red compared with Barbados coral reefs, without uplift correction, between 9 and 14 kyr B.P. in blue (Abdul et al., 2016) and green

**(Peltier and Fairbank, 2006). Data for Tahiti (Bard et al., 2010) and for Barbados (Peltier and Fairbank, 2006; Abdul et al., 2016) are only those of Acropora palmata for clarity. (B) Coral reef elevation vs age from 11.5 to 10.5 kyr before present. No uplift correction. The transition duration from the uplift rate of 0.34 mm/yr (Peltier and Fairbank, 2006; Abdul et al., 2016) to 0.8 mm/yr (Carlson and Clark, 2012; Bard et al., 2016) is of 0.2 kyr and occurred from 11.28 to 11.08 kyr BP. Data are from Peltier and Fairbank (2006) and Abdul et al. (2016).**

In the Northeast part of the island, the absence of Quaternary rocks on an area of ~60 km² suggests significant erosion or/and landsliding during the last 500-600 kyr. The volume of missing material caused by erosion and/or landslides in the Scotland District is significant (~10 km³). Furthermore, bathymetric data suggests that missing material extended offshore from the Scotland District (Figure 1B). Landslides and erosion occurred in this area during the last centuries (Prior and Ho, 1972; Cruden et al., 2014), and are documented deep offshore on the accretionary prism of Barbados (Pichot et al. 2016).

### 3 Methods

#### 3.1 Isostatic adjustment modelling

In this study, the role of abrupt mass unloading on uplift rate variation is investigated for the period from 14 kyr to 9 kyr BP by modelling the isostatic adjustment. Isostatic adjustment is considered as an additional cause of uplift and is the only process modelled here, implemented in Fortran using a finite difference numerical method. It must be added to the other mechanisms responsible for the preexisting 0.34 mm/yr of uplift rate.

Different unloaded masses were simulated corresponding to various displaced volume. Three different volumes were considered (3 km³, 10 km³, 50 km³). Topography suggests that a maximum of 10 km³ could have been displaced from the Scotland District by erosion or landslides. Offshore, erosional features observed suggest supplementary mass unloading (Figure 1B). Such volumes have been observed for a large variety of landslides or erosion. The estimation of the uplift depends also on the flexural rigidity.

In the Scotland District, the uplift $w(x)$ can be obtained by solving the equation $\nabla^2(D \cdot \nabla^2 w(x,y)) + (\rho_a - \rho_s) g \cdot w(x,y) = \rho_s g [z_{init}(x,y) - z(x,y)]$ (equation 1), where the rigidity is defined by $D = E T_e^3/[12(1-\nu^2)]$ (equation 2), with $\rho_a$ and $\rho_s$ the densities of asthenosphere and sedimentary rock, $g$ the acceleration of gravity, $z_{init}-z$ the missing material thickness, $E$ the Young's modulus, $T_e$ the effective elastic thickness, and $\nu$ the Poisson's ratio (Turcotte and Schubert, 2001). $\nu=0.25$, $\rho_a = 3179$ kg/m³, $\rho_s = 2800$ kg/m³, $E = 10^{10}$ Pa, 40 km $< T_e <$ 60 km.

The elastic thickness in the Barbados area is of 50 ± 10 km (Lambeck et al., 2014; Jimenez-Diaz et al., 2014) corresponding to a flexural rigidity of around $10^{23}$ N.m². Resulting vertical motion rates are calculated considering a constant displacement, during 10 kyr after the abrupt mass displacement, consistent with the time necessary to relax the viscous properties of the lithosphere of 10 ± 5 kyr after isostatic adjustment (Mitrovica et al., 2000; Van der Wal et al., 2010).

To avoid any overestimation of the uplift, the mass balance has been respected by loading onto the model a mass comparable to the material which has been unloaded. The loading mass is located offshore, 35 km northeast of Barbados. In case of erosion and turbidite flow, this could overestimate the loading.

### 3.2 Comparison between modelled and observed uplift

In the present study, the modelling results concerning the uplift are compared with paleo records. In Barbados, coral reefs permit to evaluate relative sea level variation and local vertical movement by comparing with a reference sea level curve. Barbados coral reef data used in this study were published by Peltier and Fairbank (2006) and Abdul et al. (2016). The sea level curve considered as a reference is based on the data from Bard et al. (1990b, 1996, 2010). Only elevation obtained from Acropora palmata and U-Th are considered. Finally, a new reference sea level curve is built using the reference sea level of Tahiti and the relative sea-level of Barbados after correction of the uplift rate. This new reference curve for sea level includes more data than the Tahiti reference curve and is still homogeneously constructed (same coral reef specie, U-Th method). Sea level trend from 14 to 9 kyr BP, based on Barbados and Tahiti coral reef records corrected from uplift, is described. Implications for meltwater pulse MWP-1B are discussed.

## 4 Results

Abrupt mass unloading, such as large erosion and landslide, causes an uplift through an isostatic adjustment and impacted relative sea level variation. More precisely, the modelled uplift is less than 2 m for an unloading of $8.4 \times 10^{12}$ kg, corresponding to a volume displacement of 3 $km^3$ of sedimentary rocks (Figure 4). The modelled uplift after an abrupt mass unloading of 10 $km^3$ is of $3.25 \pm 0.75$ m, depending on the elastic thickness of the lithosphere (Figure 5). When the mass unloading increases, the resulting uplift also increases. An abrupt mass unloading of 50 $km^3$ could cause an uplift of $6 \pm 1$ m (Figure 6A). The more the elastic thickness is, the less the vertical displacement. The uplift caused by this process ranges from 3 to 6.5 m for a mass displacement of 20-40 $km^3$.

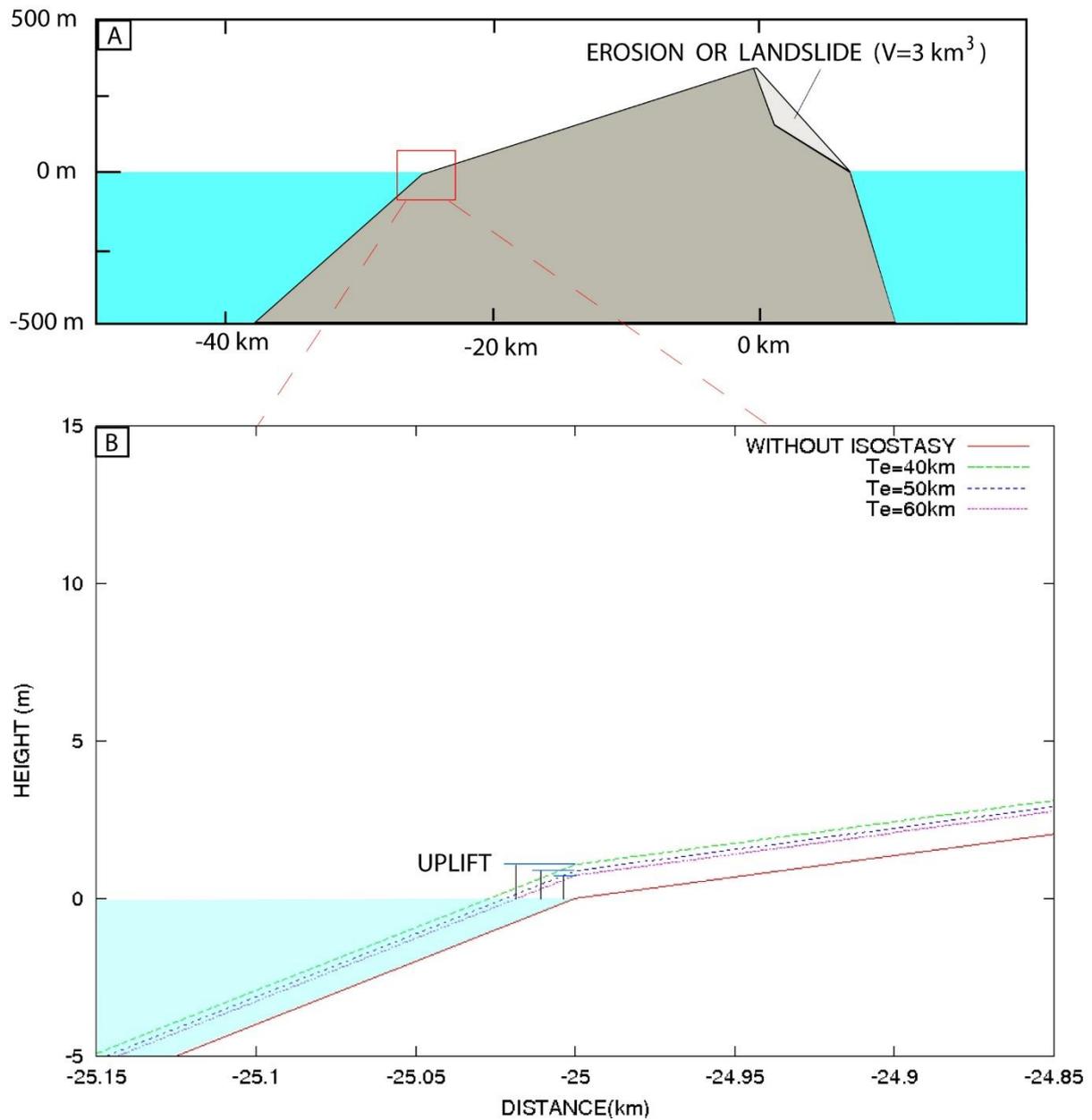

**Figure 4:** Isostatic adjustment after an abrupt mass unloading of 3 km³. (A) Unloaded volume geometry, (B) Modelled uplift in the southwest part of the island after an abrupt mass unloading caused by an isostatic adjustment with elastic thicknesses of 40 km, 50 km and 60 km.

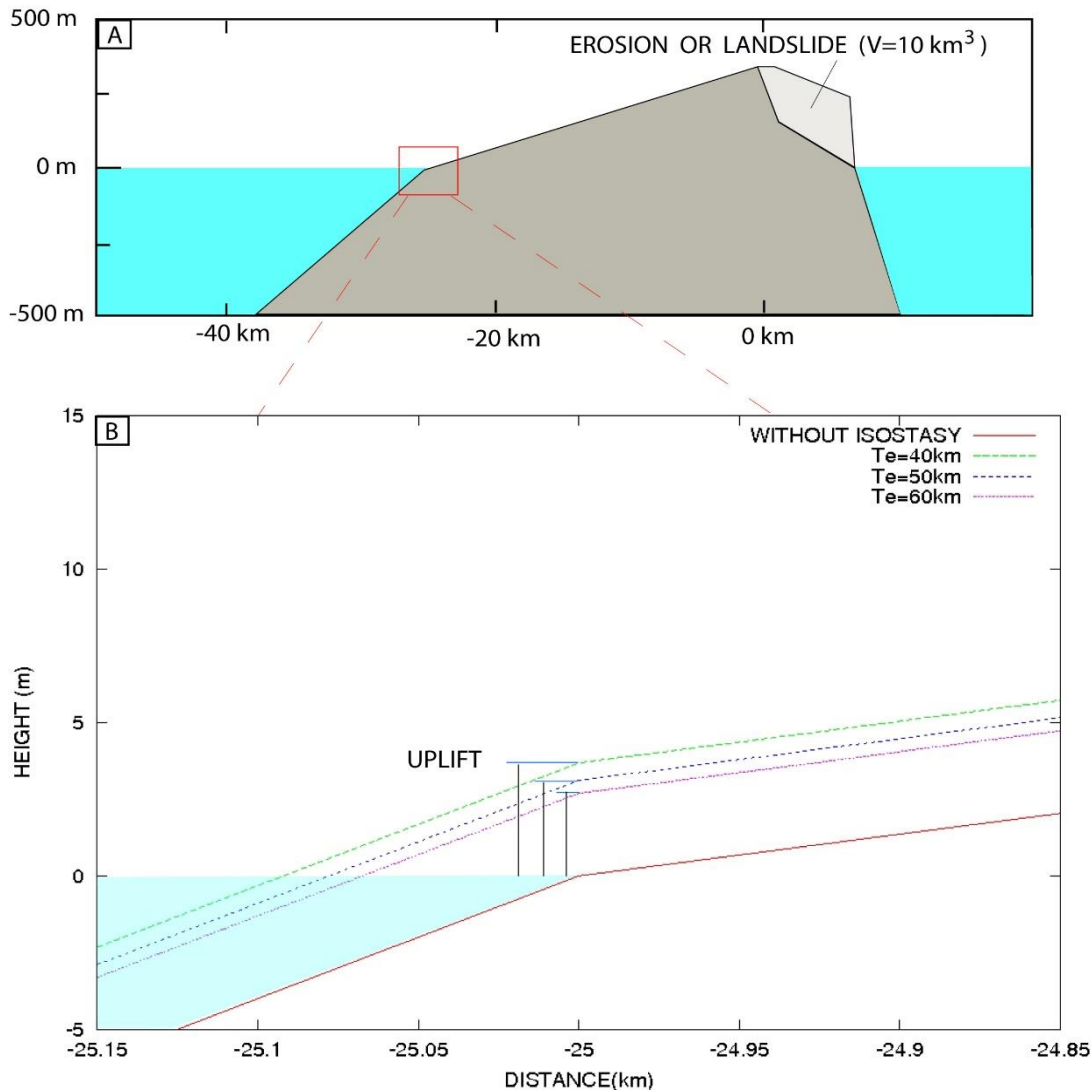

**Figure 5: Isostatic adjustment after an abrupt mass unloading of 10 km$^3$.** (A) Unloaded volume geometry, (B) Modelled uplift in the southwest part of the island after an abrupt mass unloading caused by an isostatic adjustment with elastic thicknesses of 40 km, 50 km and 60 km.

An uplift rate of 0.45 mm/yr in the southwest of Barbados, caused by an abrupt mass unloading of 30 ± 10 km$^3$ in the northeast of Barbados, is estimated by modelling the Barbados lithosphere with an effective elastic thickness of 50 ± 10 km (Fig. 6C). Part of the abrupt mass unloading (20 km$^3$) occurred offshore. The increases of 0.45 ± 0.15 mm/yr after an abrupt mass unloading of 30 ± 10 km$^3$ in the Scotland District explain the uplift rate increases observed in the Southwest coast of Barbados from 0.34 to 0.8 mm/yr that occurred 11.2 kyr ago. The change in the uplift rate occurred in less than 0.2 kyr (Figure 3B). The accommodation of the uplift rate variation took place from 11.08 kyr to 11.28 kyr BP.

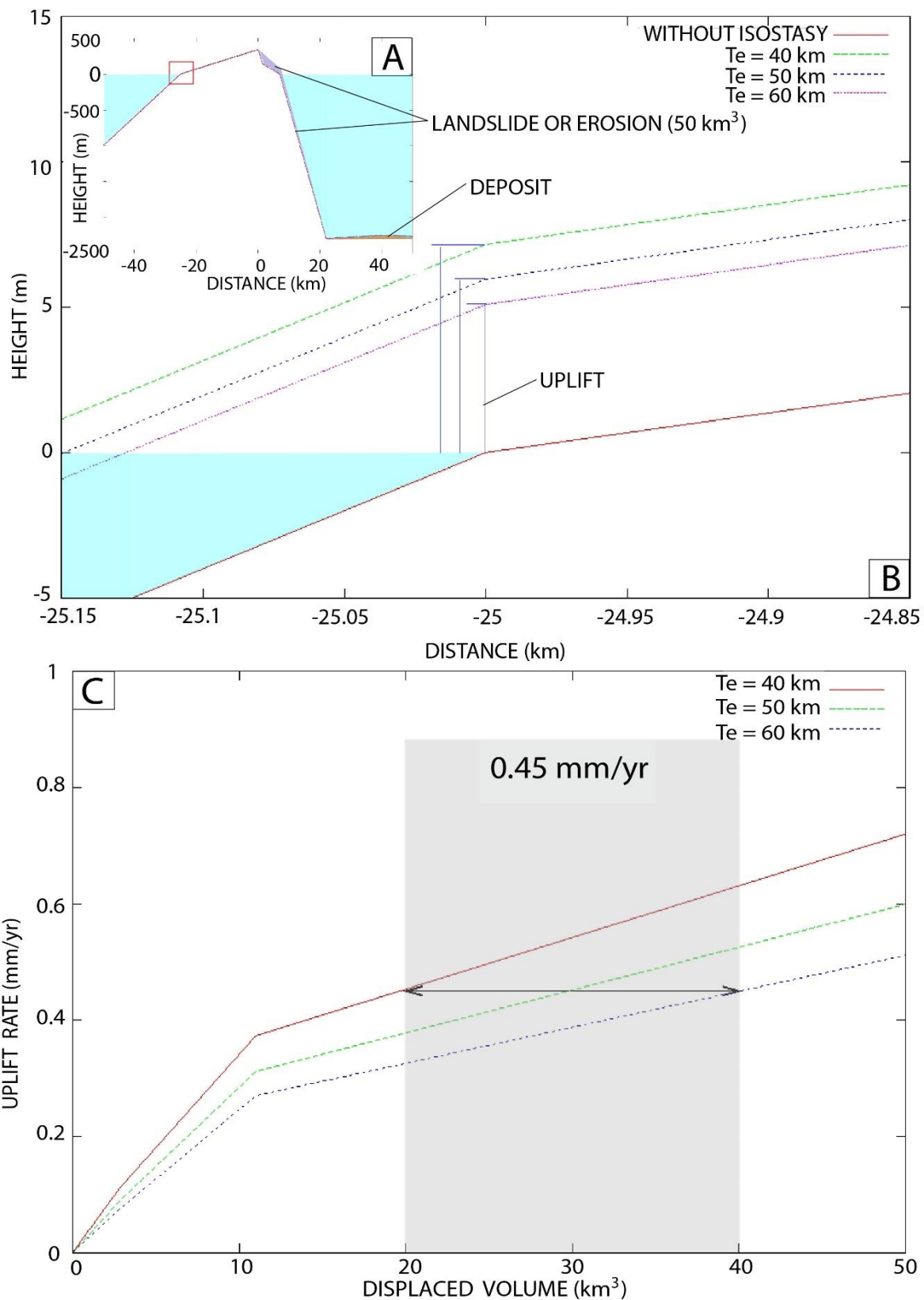

**FIGURE 6:** Isostatic adjustment modelling. (A) Topography and bathymetry modelling of Barbados Island before and after the abrupt mass unloading by erosion or landslides. The red square represents the area near Bridgetown, where coral reefs have been

**collected in previous studies at the Southwest of the Barbados and is detailed in Figure 1B. (B) Modelling of the isostatic adjustment after an abrupt mass unloading of 50 km$^3$ for different elastic thicknesses T$_e$ with values of 40, 50 and 60 km. (C) Comparison of the uplift rate caused by the isostatic adjustment triggered by abrupt mass unloading with various volumes and for different elastic thickness. An abrupt unloading by a rock volume of 20-40 km$^3$ causes an uplift rate of 0.45 mm**

This increase of uplift after 11.2 kyr permit to interpret the elevation of coral reefs from 14 kyr to 9 kyr consistent with sea-level reconstruction from other areas. After reconstruction, a rapid sea-level jump of 4.8 m is observed contemporaneously with Meltwater pulse 1B, 11.5 kyr ago, 150 yr after the end of the Younger Dryas (Fig. 7), and 300 yr before the abrupt mass unloading. The period of sea-level jump was very short (i.e. 0.1 ± 0.05 kyr). This represents a sea-level rise of 65 ± 35 mm/yr, whereas the mean sea-level rise from 14 to 9 kyr BP was of 10.4 mm/yr.

## 5 Discussion

### 5.1 Uplift rate variation with time

During the last thousand years, an uplift rate of 0.34 mm/yr was considered by previous studies (Peltier and Fairbank, 2006; Abdul et al., 2016). From 14 to 11.2 kyr BP, the 0.34 mm/yr of uplift rate were caused by tectonic compression or diapirism (Speed, 1981, Westerwood et al, 1988; Deville et al., 2006). However, from 11.2 kyr to 9 kyr, an uplift rate of 0.8 mm/yr explain the observed data, and is consistent with sea level variations used as a reference. Sea level reference curve could be constructed using a tectonically stable area where coral samples have been collected following a homogeneous protocol and accurately dated by U/Th method (Bard et al., 2010) or measuring the mean curve of numerous studies using heterogeneous methods but with an accurate statistical approach (Lambeck et al., 2014). If these curves converge for the last thousand years, they differ for older ages and the last glacial period (>14.5 kyr BP). In this study, the curve obtained in Tahiti Island (French Polynesia) using Bard et al. (2010) data was considered as a reference.

This uplift was partly (i.e. 0.45 mm/yr ; 56%) caused by the isostatic adjustment related to an abrupt mass unloading in the northeast of Barbados. Unloading associated with large landslides is able to generate isostatic adjustment of more than 10 m (Smith and Wessel, 2000). High amount of erosion also generates significant isostatic adjustment (Gargani, 2004) of which the main observed features are uplift and geomorphological transformations of the slope (Gargani et al., 2010). Nevertheless, the duration of the uplift rate increase is limited. Isostatic adjustment could explain the uplift increase during a short period of time (i.e. 10 kyr). A non-linear dynamic in vertical movements could be caused by viscoelastic rheology (Turcotte and Schubert, 2001; Gargani et al., 2006b) but has not been taken into account in this study. Uplift changes during a longer period (>> 10 kyr) cannot be explained by this process. It is possible that such process may have played a role before the 11.2 kyr event, but this is beyond the scope of this study.

Previous studies, considered that these different uplift rates were representative of two distinct areas separated by a tectonic structure (Carlson and Clark, 2012; Bard et al., 2016). Consequently, the elevation of coral reefs of core 7, 13 and 15 in the southwest of Barbados were questioned (Carlson and Clark, 2012) and Barbados sea-level reconstruction after 11.5 kyr BP was considered as not representative (Bard et al., 2016). Nevertheless, due to the older age of coral reef located on core 7, 13 and 15 in comparison to the other data, it can be also considered that a temporal variation of vertical movement occurred as in Tahiti. Barbados coral reefs can be used, as those of Tahiti, to discuss sea-level variation during the last 14 kyr BP considering appropriate uplift rates. An uplift rate of 0.8 mm/yr after 11.2 kyr BP provide coherent results with previous sea-level reconstruction (Bard et al., 2010) and previous uplift estimates (Carlson and Clark, 2012; Bard et al., 2016) in Tahiti. The cause of this uplift change is now explained.

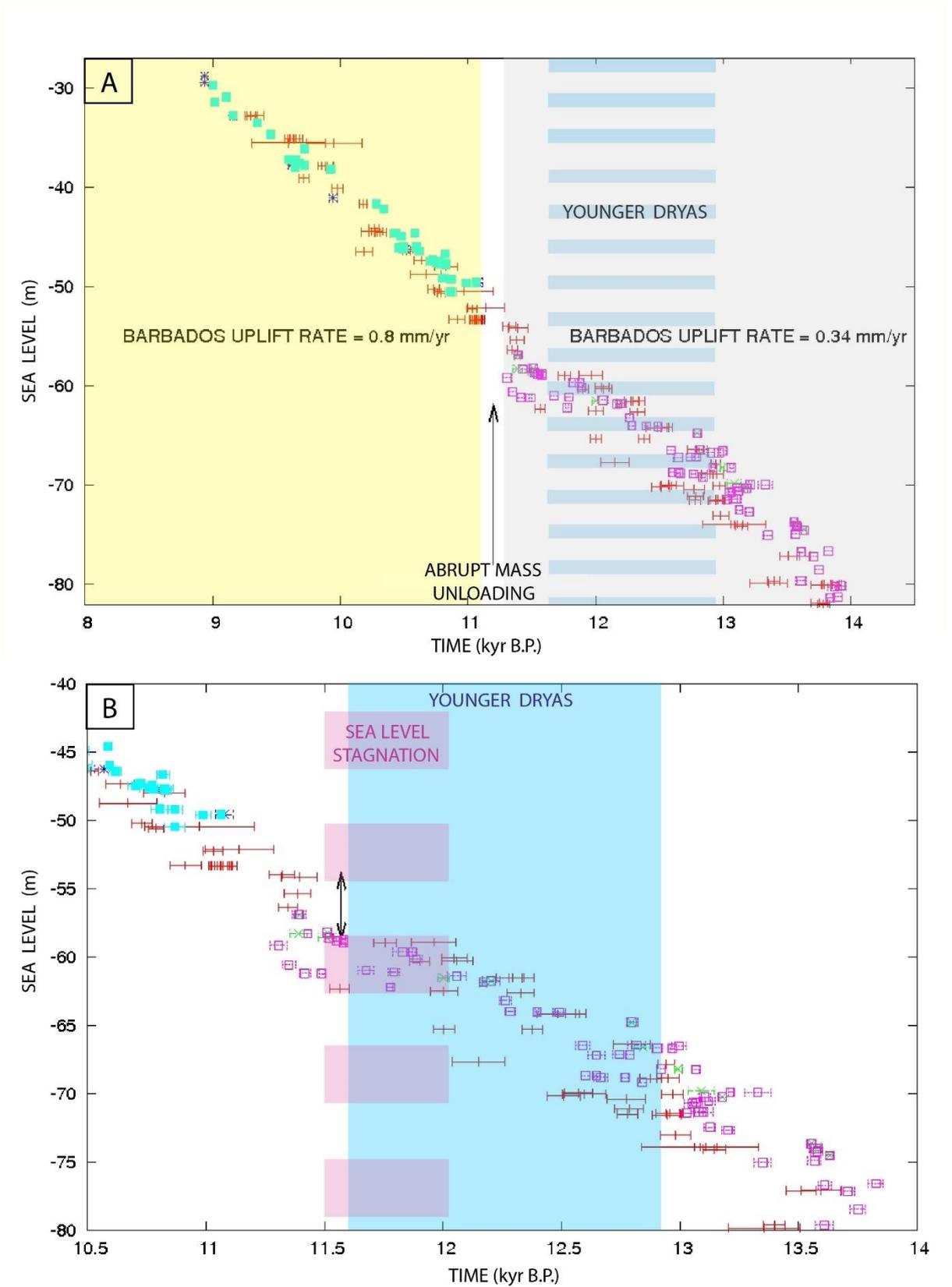

**FIGURE 7:** Sea level variation with time. (A) Sea Level at different ages from Tahiti coral reefs data in red (Bar et al., 2010) corrected from a subsidence of 0.25 mm/yr and Barbados coral reefs data (Peltier and Fairbank, 2006; Abdul et al., 2016) with a correction of 0.8 mm/yr after 11.2 kyr in light blue (Abdul et al., 2016) and dark blue

**(Peltier and Fairbank, 2006) and 0.34 mm/yr before 11.2 kyr in green (Peltier and Fairbank, 2006) and pink (Abdul et al., 2016). (B) Sea-level reconstruction during the period 13.5 to 10.5 kyr from coral reef with correction of uplift for Barbados (0.8 mm/yr after 11.2 kyr and 0.34 mm/yr before 11.2 kyr) and subsidence for Tahiti (0.25 mm/yr). Data from Peltier and Fairbank (2006) in green and dark blue, Bard et al. (2010) in red, Abdul et al. (2016) in pink and light blue.**

5.2 Sea-level variation and climatic change

Sea-level rise from 14 to 9 kyr BP is almost constant at 10.4 mm/yr considering these corrections (Figure 7), in line with previous studies in Tahiti, Barbados and Huon (Bard et al., 2010). From 14 to 12.5 kyr BP, Lambeck et al. (2014) found that the rate of sea-level rise was near 12 mm/yr, similar to that observed at Barbados and Tahiti, but also at Sunda, Huon Peninsula, Australia, New Zealand, the Indian Ocean and the Yellow and East China seas. The main evolution during this period is between 12.0 and 11.5 kyr BP when a stagnation of sea-level occurred at −60.5 ± 1.5 m. This sea-level perturbation is due to the Younger Dryas cold event estimated to occur from the onset 12870 ± 30 yr BP to its termination 11610 ± 40 yr BP (Cheng et al., 2020). At 11.5 kyr BP, around 150 yr after the end of Younger Dryas, the present study shows that a rapid sea-level jump of 4.8 m occurred in less than 0.1 ± 0.05 kyr representing a sea-level rise of 65 ± 35 mm/yr, faster than previously estimated (i.e. 40 mm/yr in Abdul et al., 2016; 16.5 mm/yr in Lambeck et al., 2014), contemporaneously with mid-Atlantic discharge increase (Tarasoc and Peltier, 2005). This sea-level jump is smaller than previously believed (i.e. 4.8 ± 1 m instead of value ranging from 7.5 ± 2.5 m to 14 ± 2 m; Fairbank, 1989; Blanchon and Shaw, 1995; Abdul et al., 2016) and occurred earlier than previous estimates (i.e. at 11.5 kyr BP instead of 11.45-11.1 kyr BP; Abdul et al., 2016), supported also by Lambeck et al. (2014). This sea-level jump is not an artifact due to tectonic and isostatic adjustment, but was caused by ice sheet melting in response to North Atlantic warming. Spatial variation of the uplift rate in Barbados between 14 and 9 kyr BP is not necessary to interpret coral reef elevation, while applying temporal variation permit to obtain coherent results between Tahiti and Barbados sea-level reconstruction. The higher uplift rate (i.e. 0.8 mm/yr) applied for correction of coral reef elevation that are older than 11.2 kyr BP causes discrepancies with the reference sea-level curve and are not justified.

As previous studies (Bard et al., 1996) have showed, there is a delay between sea level abrupt rise and Younger Dryas cold event termination that could be explained by the thermal inertia of large ice sheet and ocean dilatation. It lagged by 150 yr instead of 400 yr (Abdul et al., 2016) from Younger Dryas, but contemporaneously with meltwater megafloods from lake Agassiz (Smith et al., 1993; Blanchon and Shaw, 1995) or Lena River (Tesi et al., 2016). The main footprint of Younger Dryas on sea level is short (i.e. 0.5 kyr), but a slight reduction of sea-level rise began around 12.80 kyr BP.

5.3 Cause of abrupt mass unloading

The abrupt mass unloading occurred 0.3 kyr after the rapid sea-level jump and 0.45 kyr after the termination of Younger Dryas (Figure 8). At the termination of Younger Dryas, the climate warmed (Cheng et al., 2020) and became more humid. Sea-level increases and wetter conditions are favorable to erosion and landslide (Gargani et al., 2006; Brunetti et al., 2010; Urgeles and Camerlinghi, 2013; Gargani et al., 2014; Gargani, 2020). This event occurred a short time after the melting water pulse MWP-1B, contemporaneously with an increase of the number of landslides (Owen et al., 2007; Smith et al., 2011). If associated with erosion processes (Gargani, 2004), the abrupt mass unloading of Barbados corresponds locally to a maximum vertical erosion rate of ~1 m/yr and a regressive erosion rate of 37.5 ± 12.5 m/yr during 0.2 kyr, representing fast headcut retreats in comparison to other erosion processes (Gargani, 2004; Gargani et al., 2010; Gargani et al. 2016; Vanmaercke et al., 2016).

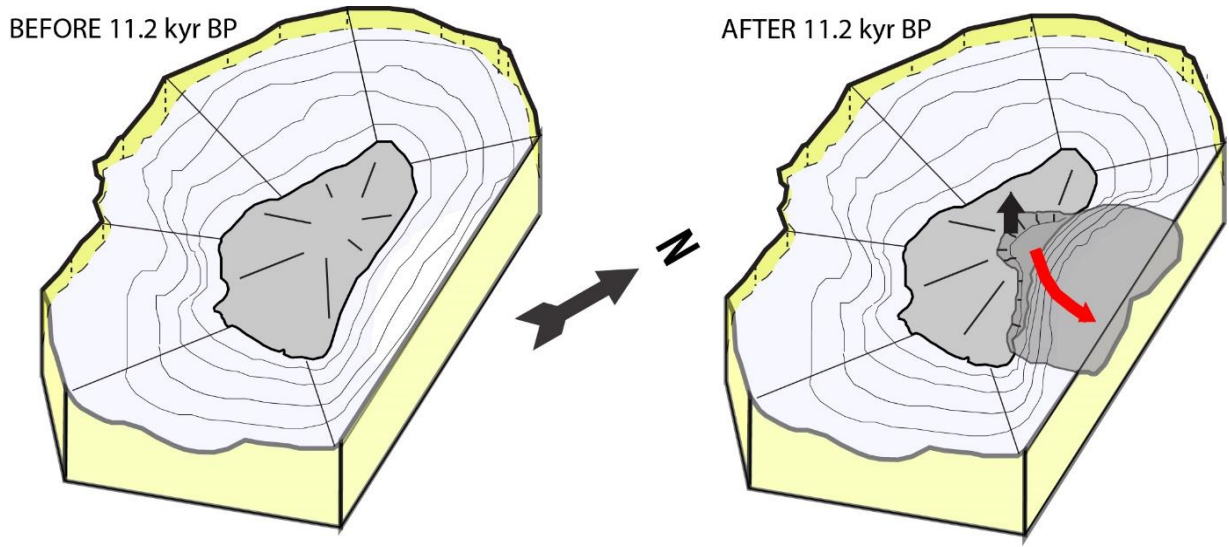

**FIGURE 8:** Schematic representation of the unloading process on geomorphology. Red arrow represents landslide or erosion process. Black arrow represents uplift.

## 6. Conclusions

Despite controversial results and continuous debate about the existence of meltwater pulse 1B in sea level, our results suggests that it is not an artefact, but that the rapid sea-level jump is smaller than previously estimated. Approximately 150 yr after the termination of Younger Drias, a sea-level jump of 4.8 m occurred, contemporaneously with megaflood events, after a short sea-level stagnation of 500 yr at −60.5 ± 1.5 m. An abrupt mass unloading occurred 300 yr after the abrupt sea-level rise. The abrupt mass unloading (landslide or/and erosion) occurred 11.2 kyr BP and caused isostatic adjustment. Isostatic adjustment, caused by an abrupt mass unloading of 30 ± 10 km$^3$ in less than 0.2 kyr, generated an uplift increase of 0.45 ± 0.15 mm/yr in Barbados. The uplift increase of 0.45 ± 0.15 mm/yr from 0.34 to 0.8 mm/yr at 11.2 ± 0.1 kyr BP explains coral reef elevation in Barbados that was not fully coherent with others sea level curves, previous to this study. Relative sea-level curves could be used to detect abrupt mass unloading.

**Acknowledgments:** The editor Zhongyuan Chen and two anonymous reviewers are acknowledged for their constructive comments and suggestions. Duncan Thom kindly helped to edit English language.

The author declares no competing interests.